\documentclass[conference,a4paper]{IEEEtran}
\IEEEoverridecommandlockouts
\usepackage{cite}
\usepackage{amsmath,amssymb,amsfonts}
\usepackage{algorithmic}
\usepackage{subfigure}
\usepackage{graphicx}
\usepackage{textcomp}
\usepackage{xcolor}
\usepackage{balance}

\def\BibTeX{{\rm B\kern-.05em{\sc i\kern-.025em b}\kern-.08em
    T\kern-.1667em\lower.7ex\hbox{E}\kern-.125emX}}
\begin{document}

\newcommand{\ac}[1]{{\color{cyan}[Achiel: #1]}}
\newcommand{\zc}[1]{{\color{blue}[Zach: #1]}}

\title{Evaluation of 3D Terrestrial and Aerial Spectrum Sharing with Massive MIMO Systems}
\author{Achiel Colpaert$^{1,2}$, Zhuangzhuang~Cui$^{2}$, and Sofie Pollin$^{2}$\\
$^{1}$imec, Kapeldreef 75, 3001 Leuven, Belgium\\
$^{2}$Department of Electrical Engineering (ESAT), KU Leuven, Belgium\\
Email: \texttt{achiel.colpaert@imec.be, \{zhuangzhuang.cui, sofie.pollin\}@kuleuven.be}}% <-this % stops a space
%\thanks{Manuscript received June 19, 2022; revised August 26, 2022.}
%\thanks{All authors are with the Networked Systems Group, WaveCoRE, Department of Electrical Engineering (ESAT), Katholieke Universiteit Leuven (KU Leuven), 3000 Leuven, Belgium (e-mail:~zhuangzhuang.cui@kuleuven.be).}}

\maketitle

\begin{abstract}
Connecting aerial and terrestrial users with a single base station (BS) is increasingly challenging due to the rising number of aerial users like unmanned aerial vehicles (UAVs). Traditional BSs, designed with down-tilted beams, focus mainly on ground users, but massive MIMO (mMIMO) systems can significantly enhance coverage in low-altitude airspace. This paper analyzes how a mMIMO BS serves both aerial and terrestrial users in a 3D spectrum-sharing scheme. Using Semi-orthogonal User Selection (SUS) and random scheduling, we assess the spectral efficiency and performance limits of these systems. Results reveal that mMIMO effectively supports more terrestrial users, influenced by channel characteristics and user scheduling strategies, providing key insights for future 3D aerial-terrestrial networks.
%Connecting aerial and terrestrial users with a single base station (BS) in cellular systems becomes highly demanding, given the increasing number of aerial users such as unmanned aerial vehicles (UAV). However, traditional BSs equipped with multiple antennas generally are designed with down-titled beams, serving ground users. Therefore, using a massive multi-input multi-output (mMIMO) system can significantly increase coverage in low altitudes where drone users are operated. This paper aims to analyze how a mMIMO BS can serve two groups of users below and above the BS, and the performance limits of such a three-dimensional (3D) spectrum-sharing scheme can achieve. We employ..., the finding..., the application. 
\end{abstract}


\section{Introduction}

Wireless networks are envisioned to serve both aerial and ground users. Emerging technologies like unmanned aerial vehicles (UAVs), drones, and other airborne platforms have become integral components of advanced communication systems, enabling use cases such as environmental monitoring, disaster management, and logistics \cite{UAV_application}. Unlike traditional terrestrial communication, where users are limited to fixed or ground-level locations, aerial users introduce a third dimension that alters the dynamics of spectrum allocation and resource management. Addressing these challenges requires innovative approaches that can flexibly allocate resources to both aerial and terrestrial users, ensuring seamless connectivity and efficient utilization of the radio frequency spectrum. Current state-of-the-art solutions, such as those proposed in \cite{bsuptilt}, consider both down-titled and up-tilted base stations (BSs). Still, there remains a gap in analyzing multi-antenna systems to handle complex interference and mobility posed by aerial users.

Massive Multiple-Input Multiple-Output (mMIMO) technology has emerged as an enabler in wireless communication, particularly for enabling efficient three-dimensional (3D) spectrum sharing. By employing a large array of antennas at the base station, mMIMO systems can perform highly directional beamforming, precisely targeting specific users while mitigating interference to others \cite{mimoofdm}. This ability to focus signal energy in desired directions is especially beneficial in scenarios involving both aerial and ground users, as it allows for dynamic adaptation to changes in user positions and altitudes. Recent advances in beamforming techniques, such as hybrid beamforming and adaptive beam steering, have further enhanced the potential of mMIMO to support diverse user distributions in 3D space \cite{mimoccuav}. Despite these advances, the integration of mMIMO with joint terrestrial and aerial communication systems requires meticulous analysis. %\ac{This study aims to bridge this gap by investigating how channel state information (CSI) can be effectively utilized to optimize beamforming patterns, thereby improving spectral efficiency for both aerial and ground users.}

Optimally scheduling the BS resources to serve both aerial and ground users is another critical challenge. Traditional scheduling algorithms, which primarily focus on ground-based users, are not well-suited to handle the dynamic propagation characteristics introduced by aerial platforms. In the state-of-the-art, researchers have proposed various strategies for user scheduling in mMIMO systems, ranging from user clustering techniques \cite{mimo_mag} to machine learning-based approaches that predict user mobility patterns \cite{mlschedule}. However, these methods often lack the flexibility to adapt to rapidly changing aerial environments, and channel state information (CSI) used for evaluation is based on simulation. % \ac{In this paper, we propose an adaptive scheduling framework} that leverages CSI at different heights to dynamically allocate resources, ensuring optimal service quality for both terrestrial and aerial users. By taking into account the specific requirements of each user type, our approach aims to maximize spectral efficiency while minimizing interference across the network.

In this paper, we first conduct CSI measurement using a mMIMO system and drone. To mimic a co-existence scenario, we flew a drone below and above the mMIMO BS. The main contributions of this paper can be summarized as follows: (i) We measure CSI for users at different altitudes, showing the channel dynamics. (ii) We demonstrate 3D spectrum sharing with mMIMO systems through advanced combining techniques. (iii) We evaluate user scheduling algorithms to investigate Spectral Efficiency (SE) and compare optimal numbers of aerial and terrestrial users. %for the base station that accounts for both aerial and terrestrial user dynamics, thereby enhancing overall network performance. %(iv) We validate our proposed methodologies through extensive simulations, showcasing the effectiveness of our approach in improving spectrum utilization and reducing interference in joint terrestrial and aerial networks.

The remainder of this paper is organized as follows: Section II introduces measurement campaigns, including system setup, measurement scenarios, and CSI data acquisition. Section III describes the user scheduling methods. In Section IV, we present the SE results in different considerations. Finally, Section V concludes the paper.
%\zc{highlighted parts should be updated.}
\section{CSI Measurement}
The mMIMO testbed operates with a frame schedule where the BS provides synchronization signaling to the users. The BS performs channel estimation to conduct MIMO signal processing, such as precoding and combining. However, in the measurement system, we focus on capturing the channel estimates, i.e., CSI, which allows for characterizing the propagation channel.

%The detailed method the BS uses to perform the channel estimation is as follows. First, the multi-antenna station broadcasts a synchronization signal to all the users. The user detects this synchronization and aligns its wavelength frame to it. Next, the user transmits a user-specific uplink pilot sequence in the designated time slots. The BS will capture these uplink pilots and perform channel estimation based on the least-square (LS) algorithm for all users simultaneously. The system then uses the obtained CSI for precoding and combining.
\subsection{Measurement System} 
We briefly describe the measurement system in terms of its setup, synchronization, and channel estimation process. The setup is the same as in our previous work, thus for more details we refer to \cite{tvtpaper}.

\subsubsection{Setup} We have the UE mounted on a DJI Inspire 2 drone. A new UE was developed using a lightweight standalone USRP E320. This device contains a Xilinx Zynq-7045 SoC together with an AD9361 RFIC. It is capable of implementing the necessary Over-the-Air (OTA) synchronization and the transmission of the uplink pilots. The built-in ARM cores of the ZYNQ processor allow an embedded Linux to automatically configure and start the software and hardware on boot. The E320 features a built-in GPSDO allowing it to synchronize its internal clock oscillator to the same GPS clock as the BS.

Besides, a low noise amplifier (LNA), i.e., Qorvo QPA9807EVB-01, is connected to the receiving antenna port. A power amplifier (PA), i.e., Qorvo TQP9111-PCB2600, is connected to the transmission antenna port to increase the dynamic range of the measurement system. Both the USRP and PAs are powered by a LiPo battery, which allows powering the devices for a longer time than the maximum drone flight duration. Finally, the antennas used at the UE are dipole antennas. 

%The drone used in the measurements is a DJI~Inspire~2. A custom platform was designed to fit all the hardware. A custom-designed mounting bracket allowed the platform to be attached to the Inspire~2. The total custom payload consists of a box containing the E320 USRP, a battery, a GPS antenna, a PA, an LNA, and 2 dipoles. The dipoles were pointed to the ground during the measurements.

%\subsubsection{Software}
The testbed is a full-scale 8-by-8 large-scale mMIMO SDR testbed, operating a total of 64 patch antennas simultaneously. The testbed runs the massive MIMO framework in Labview NX 4.0 provided by NI. It is a fully customisable framework that allows modification of both the FPGA code and host-side code. The PSS and the Uplink Pilots are the main frame symbols of interest. The PSS is used for OTA synchronization in the downlink, and the uplink pilots are used to estimate the uplink channel.

%To support channel measurements with a mobile user two main functions need to be implemented, first, the mobile user needs to synchronize its frame schedule to the BS and needs to send uplink pilots in the appropriate time slots, and second, the channel estimations in the downlink on the BS side need to be captured and written to a file.

\subsubsection{Channel Capture}
 The testbed performs channel estimation resulting in a $\textbf{H}(t)$ channel frequency response vector of dimensions $64\times1$ for each timestep $t$. Every element of the vector is a 16-bit fixed point complex number. The channel estimations are saved to DRAM during the measurement and afterwards written to a binary file. %The main limitation to the maximum measurement duration is thus the size of DRAM storage, however, the measurement duration can be extended by subsampling channel estimations written to DRAM, for example by saving a channel estimate every 1~ms the maximum measurement duration is effectively doubled.

%- labview code modification
%- frame schedule
%- LTE pilots

%\begin{figure*}
%	\centering
%    \includegraphics[scale=0.3]{img/rfnoc.jpg}
%	\caption{I think it can be changed to a table for system and setup here}
%	\label{fig:rfnoc}
%\end{figure*}

\begin{figure}
    \centering 
    \includegraphics[width=0.95\linewidth] {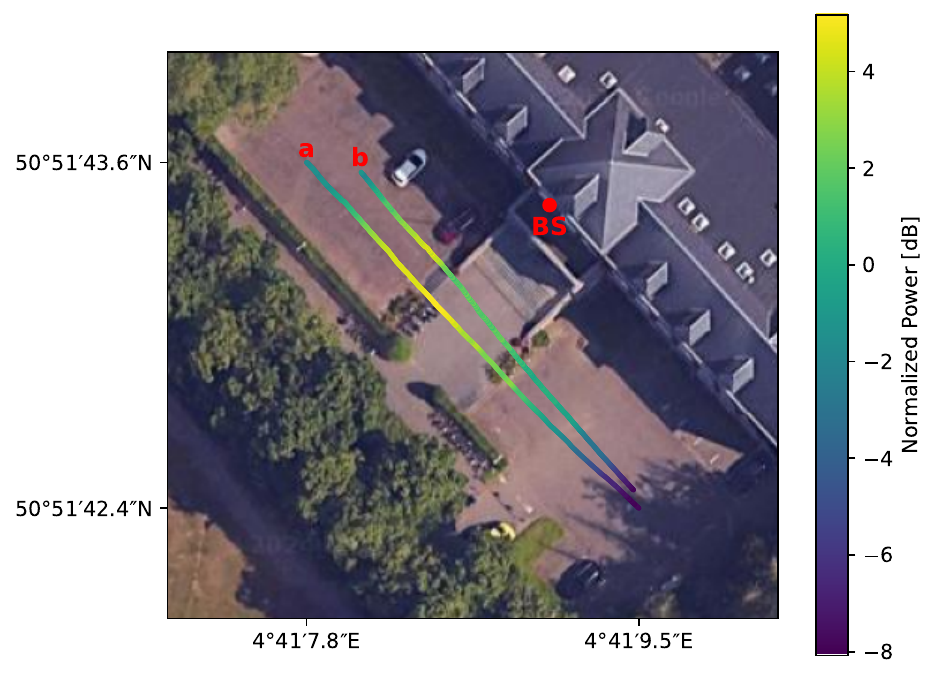}
    \caption{Measurement scenario marked by the location of BS and two trajectories that are indicated by normalized receive power. }
    \label{fig:trajectory}
\end{figure}
%\section{Measurement Setup}

\subsection{Measurement Scenario}
The target of the measurement scenario is to emulate a drone flying at different altitudes above ground level in a straight line while being connected to a BS on the ground. To emulate this, the antenna array of the BS is put in a window facing outwards at an altitude of $11~m$. The UAV flies following a trajectory parallel to the BS antenna array. The same trajectory (length: $42.48~m$) is repeated at different altitudes: $8~m$ and $24~m$, corresponding to trajectory $a$ and $b$ in Fig.~\ref{fig:trajectory}. The center frequency is $2.61~GHz$, the bandwidth is 18~MHz, the receive gain at the BS is fixed to 15~dB and the sampling interval is $\Delta t=1~ms$. During each measurement, the GPS location of the UAV was logged every $10~ms$. The drone flew at a constant speed of $1.5~m/s$ in a straight line. All of the sampling locations were in the line-of-sight (LOS) of the antenna array of the BS. Both the BS and the user used a GPS-disciplined oscillator as their input reference clock to minimize frequency offset. For more details, we refer to \cite{tvtpaper}. In the following processing, we consider every captured CSI at timestep $t$ to represent a potential UAV location and its respective CSI.

%Length trajectory 42,48m

\begin{figure}
    \centering 
      \subfigure[Summed SE versus total number of users.]{
     \includegraphics[width=0.95\linewidth]{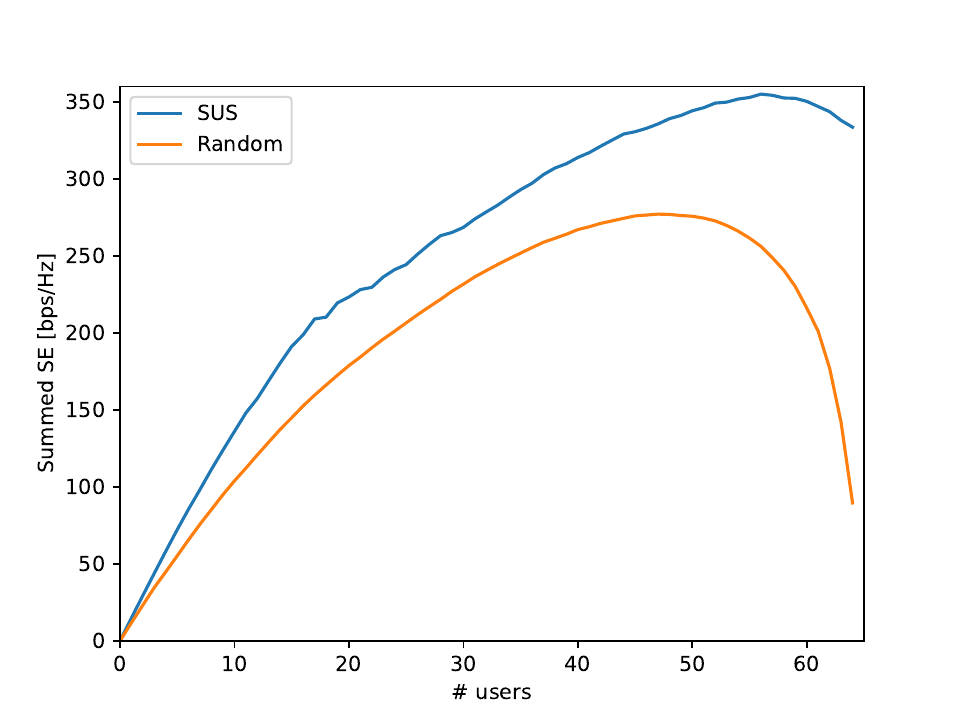}}
  \centering
  \subfigure[Individual SE versus total number of users.]{
      \includegraphics[width=0.95\linewidth]{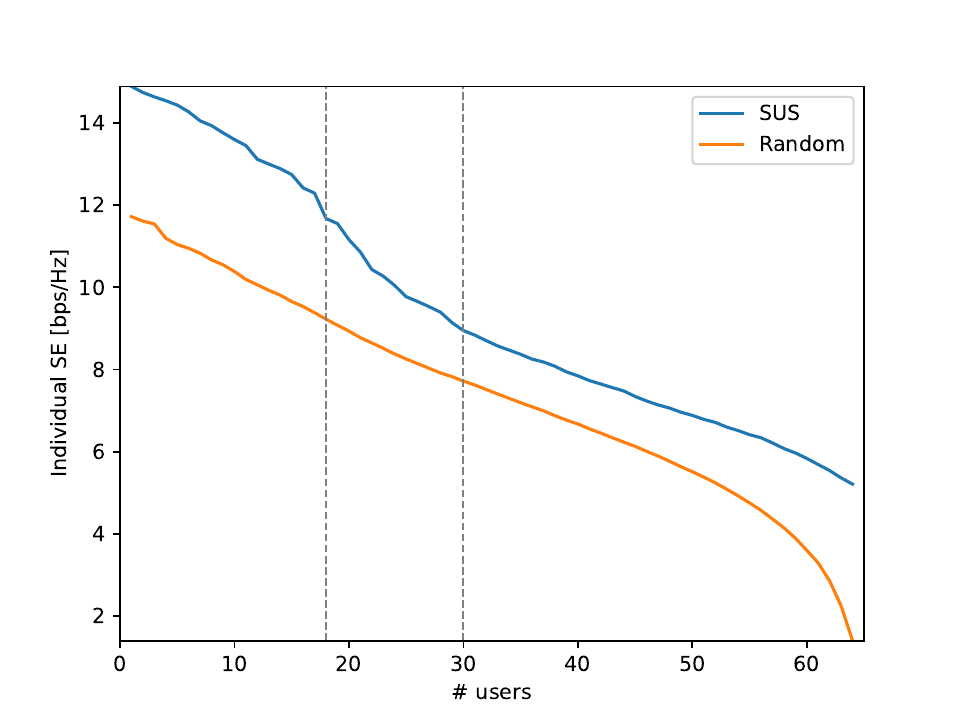}}
  \centering
    \caption{SE for different numbers of total users in the system. Figure (a) shows the summed SE and figure (b) shows the averaged individual SE for the number of users ranging from 0 to 64 where we indicate three regions with different behavior of SE.}
    \label{fig:SE}
\end{figure}

\section{Scheduling Algorithms in mMIMO with Zero-Forcing}

% Efficient user scheduling in Massive Multiple-Input Multiple-Output (MIMO) systems plays a critical role in enhancing spectral efficiency and minimizing inter-user interference. To achieve these goals, scheduling algorithms determine the optimal set of users to be served by the base station (BS) simultaneously. Two commonly utilized user selection methods in Massive MIMO systems are Semi-orthogonal User Selection (SUS) and random user selection. When combined with Zero-Forcing (ZF) beamforming, these methods can significantly improve the quality of service by directing signal power toward the intended users while suppressing interference.

Using the measured CSI, the mMIMO BS can determine the optimal set of users to schedule simultaneously. Two user selection methods are commonly used namely the random user selection and Semi-orthogonal User Selection (SUS) \cite{Yoo2006Optimality}. 

\subsection{Random User Selection}
Random user selection involves randomly choosing a set of users without considering their channel orthogonality or signal strength. While this method does not explicitly exploit the channel characteristics, it is computationally simple and can still achieve reasonable performance with ZF beamforming in scenarios where there is a high density of users. The effectiveness of ZF beamforming in this user selection method relies on the probability of achieving a sufficiently orthogonal set of users purely by chance. Random user selection will be used as a benchmark of what the average performance in a joint terrestrial and aerial communication system would be.

\subsection{Semi-orgthogonal User Selection}
The Semi-orthogonal User Selection method takes advantage of ZF beamforming by choosing a set of users whose channel vectors are as orthogonal as possible. This selection strategy is highly compatible with ZF beamforming because it inherently reduces interference when the selected users' channels are close to orthogonal \cite{Yoo2006Optimality}. The main steps in SUS are as follows:

\begin{itemize}
    \item \textbf{Candidate Selection:} Candidate user channels are projected on a semi-orthogonal basis, and the user with the largest projected norm is selected.
    \item \textbf{Basis expansion:} The newly selected user's basis projection vector is added to the existing semi-orthogonal basis.
    \item \textbf{Orthogonality Criterion Check:} From the remaining set of candidates, users are dropped based on an orthogonality condition to the new basis.
\end{itemize}

The selected user channels become semi-orthogonal with high gains, and to effectively use these semi-orthogonal channels, Zero-Forcing precoding and combining can be applied by the base station. 

Both scheduling algorithms are given a number of users to schedule and will select users from the complete dataset of locations. The remaining user locations are left out of the evaluation. This results in Random User Selection placing users in random locations in the evaluated space and in SUS placing users in the optimal spatial distribution in the evaluated space if using Zero-Forcing precoding.

\subsection{Zero Forcing}

Zero-Forcing (ZF) beamforming is a linear precoding and combining technique that aims to nullify inter-user interference by ensuring that the beamforming vectors are orthogonal to the channel vectors of non-targeted users. In mMIMO systems, ZF beamforming is particularly effective because it leverages the large number of antennas to create precise beam patterns that nullify signal power towards other users. This approach enables the system to minimize interference, thus increasing the SINR and overall spectral efficiency.

Consider a mMIMO system where the base station has $M$ antennas and serves $K$ users simultaneously. Let the channel matrix be represented as $\mathbf{H} = [\mathbf{h}_1, \mathbf{h}_2, \dots, \mathbf{h}_K]^T$, where $\mathbf{h}_k$ is the channel vector for the $k$-th user, of size $M \times 1$. We assume the base station performs perfect channel estimation, that the channel matrix is normalized and that the average Signal-to-Noise-Ratio over the whole set of CSI samples is 20~dB.

The zero forcing combining vector $\mathbf{v}_k$ for the $k$-th user is calculated as follows:

\begin{equation}
    \mathbf{V}^{ZF} = \mathbf{H}((\mathbf{H})^H\mathbf{H})^{-1}.
\end{equation}

where $\mathbf{V}^{ZF} = [\mathbf{v}_1,\mathbf{v}_2,\dots,\mathbf{v}_K]^T$, $\mathbf{H}^H$ is the Hermitian (conjugate transpose) of the channel matrix $\mathbf{H}$ and, $\left( \mathbf{H}^H \mathbf{H} \right)^{-1}$ is the inverse of the Gram matrix.

This formulation guarantees that the ZF beamforming vectors effectively minimize interference to non-targeted users, which maximizes the system's SINR and SE.

\section{Evaluation Metrics}

\subsection{SINR}

When applying ZF combining in mMIMO systems, the SINR and SE formulations need to account for the interference suppression provided by the combining vectors.

The SINR for the $k$-th user and $l$-th frequency bin when using ZF combining can be expressed as \cite{massivemimobook}:
\begin{equation}
    SINR_k = \frac{p_k|\mathbf{v}_k^H\mathbf{h}_k|^2}{\sum_{i\neq k}^{K}p_k|\mathbf{v}_k^H\mathbf{h}_i|^2+\mathbf{v}_k^H(\sigma^2\mathbf{I})\mathbf{v}_k},
\end{equation}
where $p_k$ is the transmit power of user $k$ and $\sigma^2$ is the noise power. For this evaluation we assume that $p_k$ is equal for all users.
%\zc{I guess your $P_k/\sigma^2$ is fixed in 20~dB?}

\subsection{Spectral Efficiency}

The SE for the $k$-th user, denoted as $SE_k$, can be calculated as follows \cite{massivemimobook}:
\begin{equation}
    SE_k = \log_2(1 + SINR_k),
\end{equation}
The total spectral efficiency of the system, when serving $K$ users, is given by the sum of the individual users' spectral efficiencies:
\begin{equation}
    SE_{sum} = \sum_{k = 1}^{K}SE_k.
    \label{eq:sum}
\end{equation}

\begin{figure}[!t]
  \centering
  \subfigure[Summed SE for the whole range of 0 to 28 aerial users and 0 to 36 ground users.]{
    \includegraphics[width=0.8\linewidth]{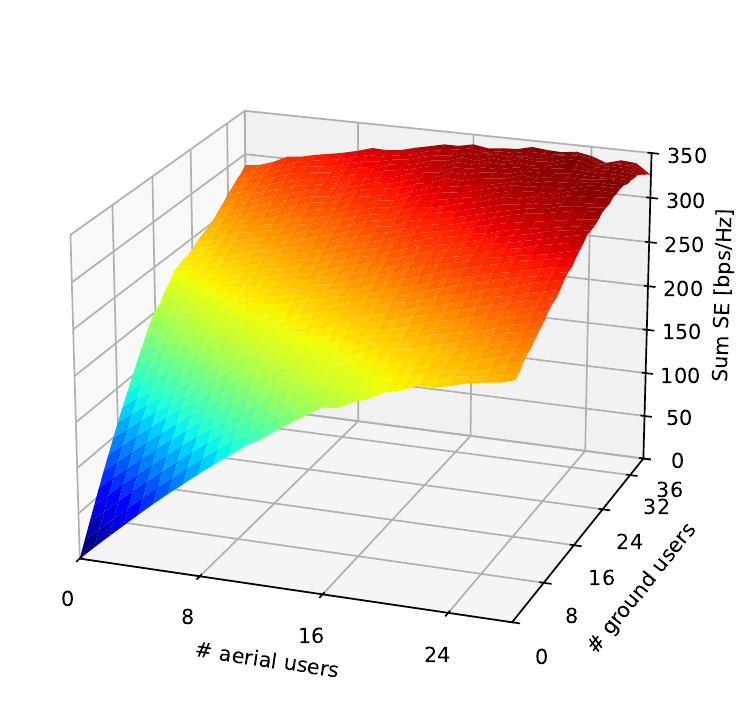}}
  \centering
  \subfigure[Zoom of Figure (a) in the region of 16 to 28 aerial users and 16 to 36 ground users.]{
    \includegraphics[width=0.8\linewidth]{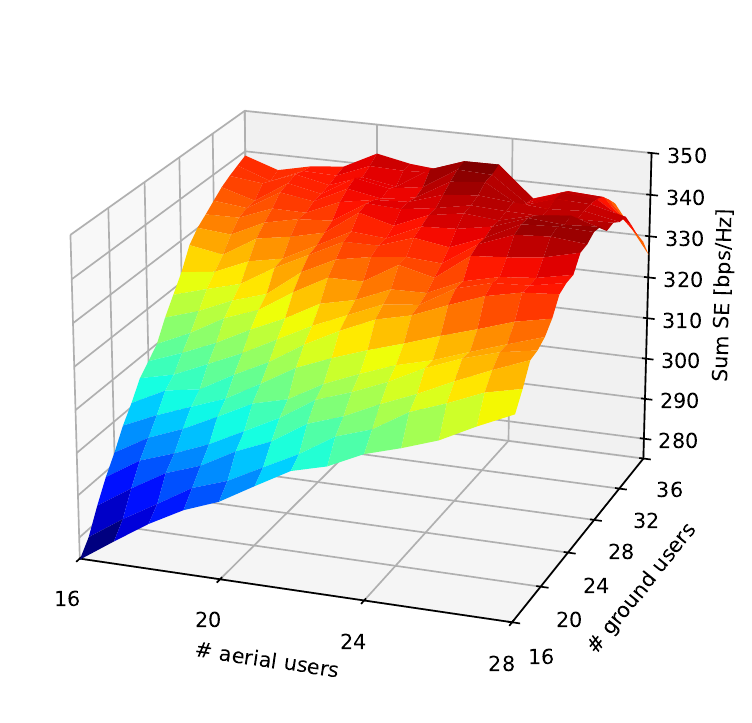}}
  \centering
  \caption{Summed SE as a function of numbers of ground and aerial users when using SUS and ZF combining.}
  \label{fig:sumSE}
\end{figure}

\section{Results}
In this section, we first show the summed SE and individual SE results with the two scheduling methods, random and semi-orthogonal user scheduling. The random scheduling acts as a benchmark where the user's channels are not taken into account when scheduling. The SUS scheduling offers the user placement in the given scenario when using ZF. We consider all the CSI samples at $24~m$ altitude, the aerial layer, and the samples at $8~m$ altitude, the terrestrial layer. Then, considering the co-existence of both layers, we investigate how to schedule users in both layers to achieve optimal SE performance. The following results don't make a distinction between aerial and terrestrial users; users get assigned either randomly in any of the layers or for the SUS method, users get assigned to the most optimal locations, ignoring whether they are on the terrestrial or the aerial layer.
\subsection{SE performance with random users in both layers}
First of all, we show the summed SE versus the total number of users in the system in Fig.~\ref{fig:SE}(a). For every number of users, the SUS and random scheduling method is applied and the achieved summed SE is plotted. It shows that both methods present the same trend: the summed SE increases first and then decreases as the total number of users approaches the number of antennas at the base station. However, SUS always achieves a better performance than when using random scheduling. We find that the highest possible summed SE in the system using SUS and ZF is $354.9~bps/Hz$ at a total number of users of 56.

For individual SE, we show the result in Fig.~\ref{fig:SE}(b). It shows a decreasing trend with an increasing number of users. Thus, we can determine the maximum number of users when the system requires a minimum average SE for each user. As an example, for average $\overline{SE}_k$ of $8~bps/Hz$, we can determine the system can maximum support 27 and 38 users for random scheduling and SUS, respectively. We can indicate three regions in the individual SE plot where the slope of the SE is different. In the center region, there is a larger penalty for individual SE when the total number of users is increased.
\begin{figure}[!t]
  \centering
    \includegraphics[width=0.8\linewidth]{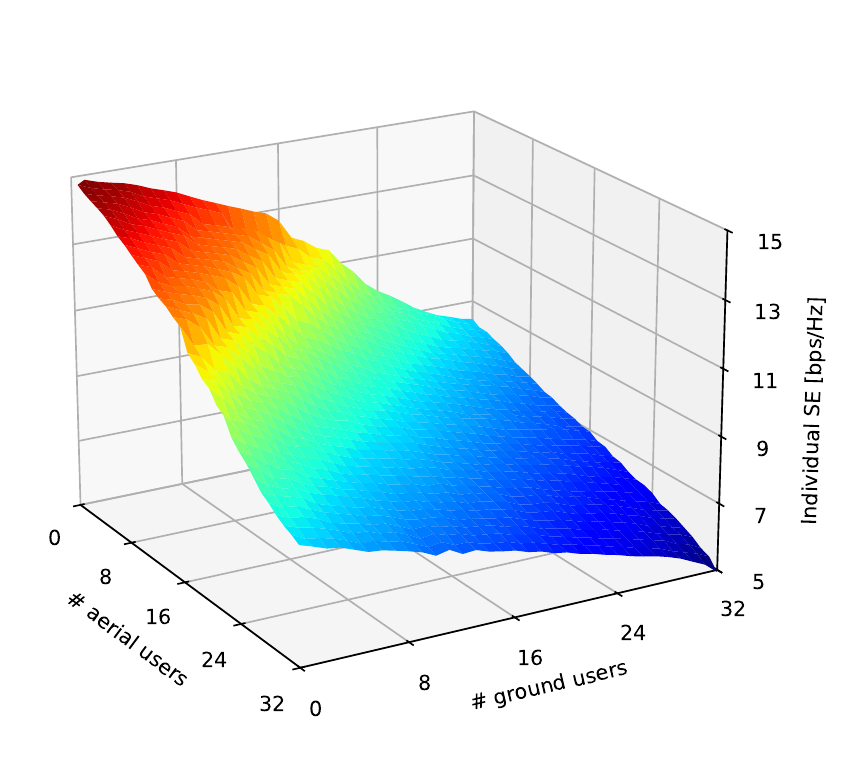}
  \centering
  \caption{Individual SE as a function of numbers of ground and aerial users.}
  \label{fig:indSE}
\end{figure}
\subsection{SE performance with designated users in each layer}
In the above evaluation, we consider selecting users in two layers randomly. However, to guarantee the co-existence of users in both layers and spectrum sharing, we consider selecting a predetermined number of users from each layer and then investigating the system's performance. 

Next, we evaluate the SUS scheduling for different combinations of numbers of users in the terrestrial and aerial layers. We achieve this by allowing the SUS algorithm to choose user locations from either layer until one of the required number of users is achieved. From that point onwards, the algorithm is only allowed to choose locations from the remaining layer.

The summed SE in function of a number of aerial users and a number of ground users can be seen in Fig.~\ref{fig:sumSE}(a). We see that the surface is almost symmetric along the diagonal, with an equal number of ground and aerial users. This indicates that for this region, there is no difference in system performance when assigning more or fewer users to the aerial layer or to the terrestrial layer. However, some irregularities can be observed at the highest summed SE in the region of around 50 to 60 total users. If we then further zoom into this region, as shown in Fig.~\ref{fig:sumSE}(b), we can see that the maximum summed SE values do not exist on the diagonal of equal ground and aerial users. We found that the maximum summed SE of $346.6~bps/Hz$ is obtained when assigning 34 users to the terrestrial layer and 24 users to the aerial layer. This can be explained by the larger stationary distance experienced by ground users when compared to aerial users. As shown in \cite{tvtpaper}, the stationary distance for these trajectories is larger than for aerial users. This means that the SUS method will have more difficulty finding orthogonal channels in the aerial layer since the correlation between the aerial user's channels is higher.

% SE reporting

%Maximum SE reporting\zc{include the optimal number from Fig. 3(b) of ground and aerial users, also maybe check whether the total number is aligning with Fig.~2(should around 50). }

% \begin{figure*}
%     \centering
%     \begin{subfigure}{0.45\textwidth}
%         \includegraphics[width=.90\textwidth]{images/cdf_sum_se_URA_lab_LoS.eps}
%         \caption{(a) URA LoS}
%         \label{fig:CDF_MR_URA}
%     \end{subfigure}
%     \begin{subfigure}{0.45\textwidth}
%         \centering
%         \includegraphics[width=.90\textwidth]{images/cdf_sum_se_ULA_lab_LoS.eps}
%         \caption{(b) ULA LoS}
%         \label{fig:CDF_MR_ULA}
%     \end{subfigure}
%    \caption{CDFs of sum SE for different user scheduling techniques (Random, SUS, ABC, and DEF) in URA LoS and ULA LoS condition using RZF as precoder.}   

%     \label{fig:SE_URA}
% \end{figure*}

%We first look at the sum SE in the 3D, as a function of numbers of ground and aerial users. Fig.~3(a) shows...

%Then, to clearly show the trend, we focus on user numbers ranging from 16 to 32, as shown in Fig.~3(b). The results demonstrate that, the sum SE decreases when the number of users becomes large \zc{I guess here aligns with FIG. 2, when number exceeds 52 (maybe), SE decreases, double check please}. So the optimal numbers of ground and aerial users should be 24, 28....
%Moving to individual SE...
Fig.~\ref{fig:indSE} shows individual SE versus numbers of ground and aerial users. It is found that it presents a symmetric behavior for varying numbers of users in both layers. Corresponding to Fig.~\ref{fig:SE}(b), we can observe the same three regions for all combinations of ground and aerial users.  Thus, considering the fairness of the given requirement of $SE_k$, we can suggest a reasonable total number of users in the network for any combination of ground and aerial users. 

\section{Conclusion}
This paper has investigated 3D spectrum sharing for aerial and terrestrial users within a mMIMO system equipped with 64 antennas. CSI was derived from real-world measurements conducted in a campus environment, considering two distinct user heights relative to the base station to represent ground and aerial user distributions. We evaluated the SE performance under SUS and random scheduling, selecting users from both layers. Our results demonstrate that a higher number of users can be supported when employing SUS. To determine the precise number of users in each layer, we conducted a 3D analysis of summed and individual SE. The findings reveal that an optimal summed SE is reached in the case of 24 aerial users and 34 terrestrial users, which appears counterintuitive at first glance. However, this outcome can be effectively explained by the combined effects of channel stationarity distance and the SUS method. Overall, this study provides valuable insights into 3D spectrum sharing with mMIMO systems and highlights the performance limits under optimal user scheduling strategies. However, the limited resolution in altitude is a limiting factor of this work, since in a real scenario UAVs are not limited to a single one or two altitudes. Using the presented measurement setup more extensive datasets and complex scenarios can be measured and evaluated providing more valuable insights into spectrum sharing between aerial and ground users to further the development of non-terrestrial communication systems.
\balance
\section*{Acknowledgment}
This work is supported by the iSEE-6G project under the Horizon Europe Research and Innovation program with Grant Agreement No. 101139291. 

\bibliographystyle{IEEEtran}
\bibliography{Reference}

\begin{thebibliography}{1}
\providecommand{\url}[1]{#1}
\csname url@samestyle\endcsname
\providecommand{\newblock}{\relax}
\providecommand{\bibinfo}[2]{#2}
\providecommand{\BIBentrySTDinterwordspacing}{\spaceskip=0pt\relax}
\providecommand{\BIBentryALTinterwordstretchfactor}{4}
\providecommand{\BIBentryALTinterwordspacing}{\spaceskip=\fontdimen2\font plus
\BIBentryALTinterwordstretchfactor\fontdimen3\font minus \fontdimen4\font\relax}
\providecommand{\BIBforeignlanguage}[2]{{%
\expandafter\ifx\csname l@#1\endcsname\relax
\typeout{** WARNING: IEEEtran.bst: No hyphenation pattern has been}%
\typeout{** loaded for the language `#1'. Using the pattern for}%
\typeout{** the default language instead.}%
\else
\language=\csname l@#1\endcsname
\fi
#2}}
\providecommand{\BIBdecl}{\relax}
\BIBdecl

\bibitem{UAV_application}
G.~Geraci, A.~Garcia-Rodriguez, M.~M. Azari, A.~Lozano, M.~Mezzavilla, S.~Chatzinotas, Y.~Chen, S.~Rangan, and M.~D. Renzo, ``What will the future of {UAV} cellular communications be? a flight from {5G} to {6G},'' \emph{IEEE Communications Surveys \& Tutorials}, vol.~24, no.~3, pp. 1304--1335, 2022.

\bibitem{bsuptilt}
L.~Chen, M.~A. Kishk, and M.-S. Alouini, ``Dedicating cellular infrastructure for aerial users: Advantages and potential impact on ground users,'' \emph{IEEE Transactions on Wireless Communications}, vol.~22, no.~4, pp. 2523--2535, 2023.

\bibitem{mimoofdm}
M.~Jiang and L.~Hanzo, ``Multiuser {MIMO-OFDM} for next-generation wireless systems,'' \emph{Proceedings of the IEEE}, vol.~95, no.~7, pp. 1430--1469, 2007.

\bibitem{mimoccuav}
Y.~Huang, Q.~Wu, R.~Lu, X.~Peng, and R.~Zhang, ``Massive {MIMO} for cellular-connected {UAV}: Challenges and promising solutions,'' \emph{IEEE Communications Magazine}, vol.~59, no.~2, pp. 84--90, 2021.

\bibitem{mimo_mag}
A.~Colpaert, S.~De~Bast, R.~Beerten, A.~P. Guevara, Z.~Cui, and S.~Pollin, ``Massive {MIMO} channel measurement data set for localization and communication,'' \emph{IEEE Communications Magazine}, vol.~61, no.~9, pp. 114--120, 2023.

\bibitem{mlschedule}
I.~M. Shawky, M.~Sadek, and H.~M. Elhennawy, ``Uplink multiuser scheduling using machine learning,'' in \emph{2020 15th International Conference on Computer Engineering and Systems (ICCES)}, 2020, pp. 1--6.

\bibitem{tvtpaper}
A.~Colpaert, Z.~Cui, E.~Vinogradov, and S.~Pollin, ``{3D} non-stationary channel measurement and analysis for {MaMIMO-UAV}communications,'' \emph{IEEE Transactions on Vehicular Technology}, vol.~73, no.~5, pp. 6061--6072, 2024.

\bibitem{Yoo2006Optimality}
T.~Yoo and A.~Goldsmith, ``On the optimality of multiantenna broadcast scheduling using zero-forcing beamforming,'' \emph{IEEE Journal on Selected Areas in Communications}, vol.~24, no.~3, pp. 528--541, 2006.

\bibitem{massivemimobook}
\BIBentryALTinterwordspacing
E.~Bj\"{o}rnson, J.~Hoydis, and L.~Sanguinetti, ``Massive {MIMO} networks: {Spectral}, energy, and hardware efficiency,'' \emph{Foundations and Trends{\textregistered} in Signal Processing}, vol.~11, no. 3-4, pp. 154--655, 2017. [Online]. Available: \url{http://dx.doi.org/10.1561/2000000093}
\BIBentrySTDinterwordspacing

\end{thebibliography}

\end{document}